\documentclass[pra,aps,twocolumn,showpacs,10pt,floatfix]{revtex4-1}
\usepackage{amsmath,amssymb,graphicx,bm,color,ulem}
\usepackage{amsfonts,amsthm}

\begin{document}

\title{Spatiotemporal optical vortex solitons: \\ Dark solitons with transverse and tilted phase line singularities}

\author{Miguel A. Porras}

\affiliation{Grupo de Sistemas Complejos, ETSIME, Universidad Polit\'ecnica de Madrid, Rios Rosas 21, 28003 Madrid, Spain}

\begin{abstract}
Vortex solitons (dark solitons) are described in self-defocusing Kerr media whose phase line singularity is not parallel to the propagation direction, but is perpendicular or tilted almost arbitrary angles, depending on the medium linear dispersion. These solitons broaden the so-called spatiotemporal optical vortices recently observed in experiments to the nonlinear realm, opening new perspectives in practical applications of optical vortices.
\end{abstract}


\maketitle

{\it Introduction.} Since a few decades, the expression spatiotemporal optical vortex in the context of solitons refers to three-dimensional solitons carrying a vortex whose phase line singularity is parallel to the propagation direction \cite{LEBLOND,DESYATNIKOV}, usually the $z$ axis in optics. The gradient of the phase, $2m\pi$, where the integer $m$ is the topological charge, circulates in a closed line about the $z$ axis in the two transverse spatial coordinates. Very recently, a new type of vortex whose phase line singularity is not parallel but transversal to the propagation direction, like in a moving tornado, has been called spatiotemporal optical vortex (STOV) because the gradient of the phase circulates in a transversal direction and time. These STOVs have first been observed to accompany the self-guided pulse after the arrest of collapse \cite{JHAJJ}, and then has been generated by linear means employing a $4f$ pulse shaper \cite{HANCOCK} or a spiral phase and a pulse shaper \cite{CHONG}. Linearly or nonlinearly generated, these wave objects behave substantially as linear waves. On a more theoretical side, spatiotemporal linear vortices with orbital angular momentum forming an arbitrary angle with respect to the momentum direction have been also described \cite{BLIOKH}.

Here we describe STOV solitons in the new sense, intrinsically nonlinear waves that are supported by self-defocusing Kerr media, with or without linear dispersion. These STOV solitons add a new, continuous degree of freedom to the standard, monochromatic optical vortex solitons (OVS) \cite{NEU,KIVSHAR}, or dark solitons, namely, the orientation of the phase line singularity, which can be tilted almost arbitrary angles with respect to the $z$ axis. This degree of freedom may broaden the perspectives of applications of vortices such as encoding information, particle trapping, and wave guiding or mixing. These STOV solitons are intrinsically polychromatic, present a $2m\pi$ circulation of the phase gradient along a closed curve about the singularity in a spatiotemporal plane, as the previously reported STOVs, and can exist in media with anomalous, vanishing and normal dispersions. With anomalous dispersion, STOV solitons with arbitrary orientation of the phase line singularity do exist, while normal and vanishing dispersion impose some restrictions. As discussed below, STOV solitons are as stable as monochromatic vortex solitons, and are expected to be formed spontaneously from linear STOVs propagating in the self-defocusing medium.

STOV solitons are described here in a paraxial regime of beam propagation of quasi-monochromatic pulses. Under these conditions, the propagation of the wave envelope $\psi(t',x,y,z)$ of the electric field $E=\psi e^{ik_0 z}e^{-i\omega_0 t}$ of carrier frequency $\omega_0$ and propagation constant $k_0$ in is ruled by the nonlinear Schr\"odinger equation (NLSE)
\begin{equation}\label{NLSE1}
\partial_z\psi = \frac{i}{2k_0} \Delta_{x,y} \psi - \frac{i k_0^{\prime\prime}}{2} \partial^2_{t'} \psi + \frac{ik_0n_2}{n_0}|\psi|^2\psi\,,
\end{equation}
where $\Delta_{x,y}$ is the two-dimensional Laplacian in the indicated variables, $n_0=n(\omega_0)$ is the refraction index at $\omega_0$, $t'=t-k'_0 z$ is the local time, $k(\omega)=n(\omega)\omega/c$ is the propagation constant, $c$ the speed of light in vacuum, $k^{(n)}_0 = d^{n}k(\omega)/d\omega^n|_{\omega_0}$, and $n_2<0$ is the negative nonlinear refractive index.

{\it Standard OVS.} We recall that monochromatic vortex solitons, or dark solitons, \cite{NEU,KIVSHAR} consist on a phase line singularity along the $z$ direction, surrounded by a cylindrically symmetric darkness on the uniformly luminous background of the nonlinear plane wave $\psi=\sqrt{I_0}e^{i\delta z}$ of intensity $I_0 =n|\delta|/k_0|n_2|$ and with $\delta<0$. It is then convenient to introduce normalized coordinates $\zeta =|\delta| z$, $(\xi,\eta)=\sqrt{k_0|\delta|}\,(x,y)$, and envelope $u=\psi/\sqrt{I_0}$, to rewrite (\ref{NLSE1}) as
\begin{equation}\label{NLSE2}
\partial_\zeta u = \frac{i}{2}\Delta_{\xi,\eta} u - \frac{ik_0^{\prime\prime}}{2|\delta|}\partial^2_{t'} u  - i|u|^2 u\,,
\end{equation}
and since the second derivative in time vanishes for $u=u(\xi,\eta,\zeta)$ independent of time,
\begin{equation}\label{NLSE2bis}
\partial_\zeta u = \frac{i}{2}\Delta_{\xi,\eta} u - i|u|^2 u\,.
\end{equation}

\begin{figure*}[!]
\begin{center}
\includegraphics*[height=4cm]{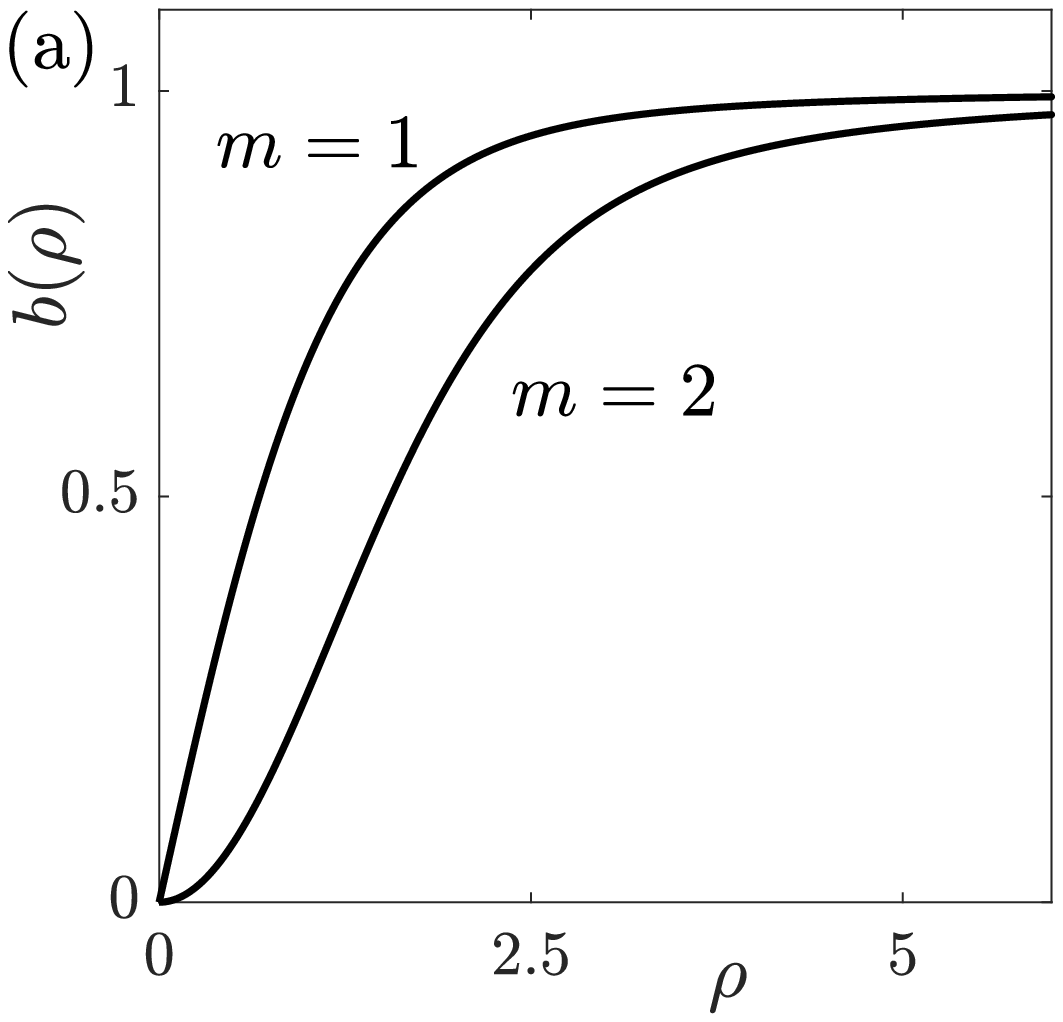}\includegraphics*[height=4cm]{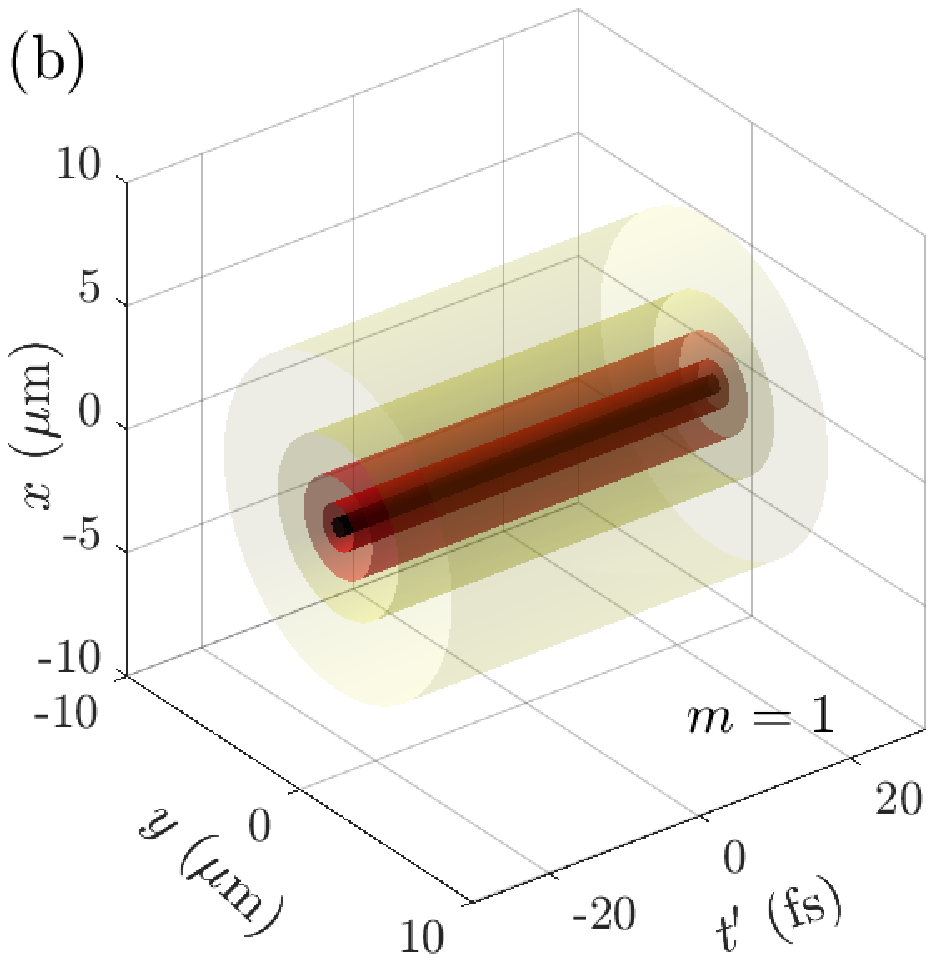}
\includegraphics*[height=4cm]{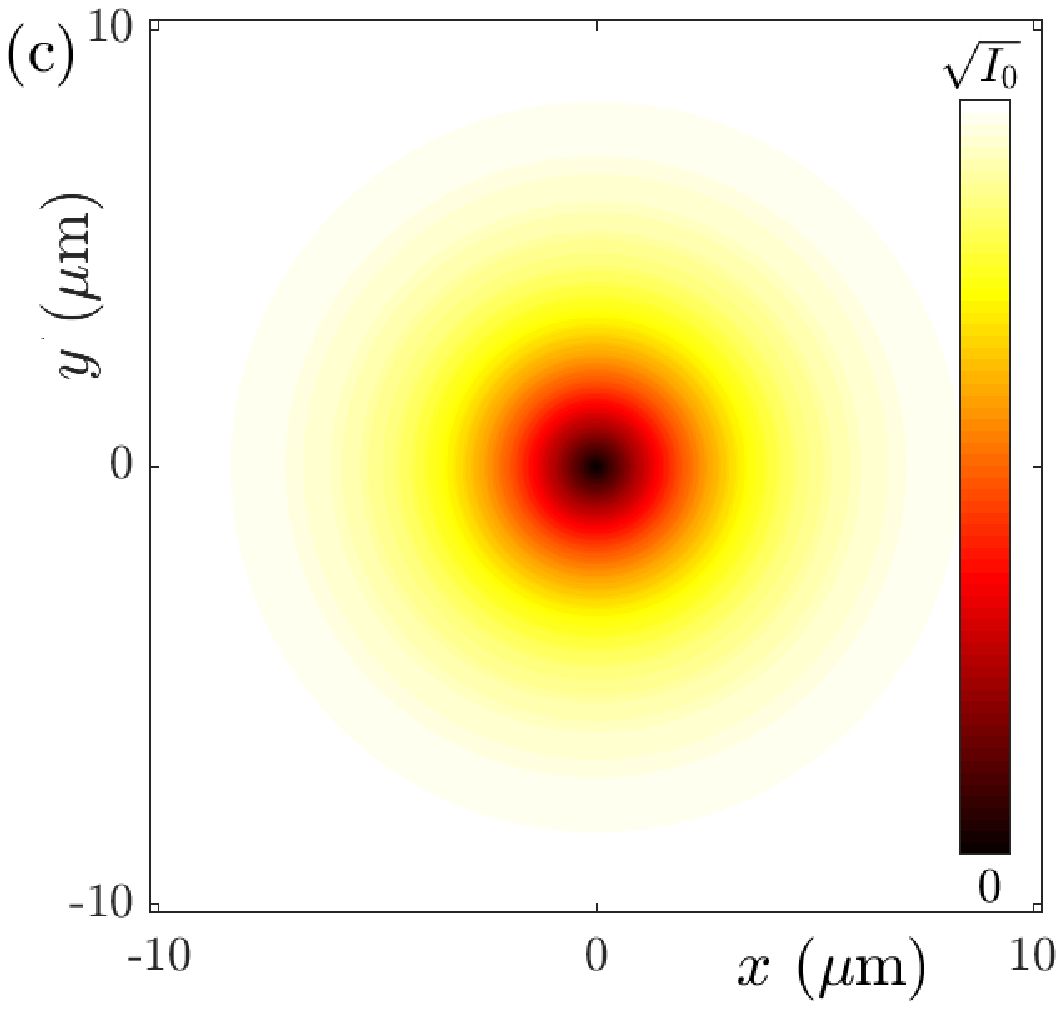}\includegraphics*[height=4cm]{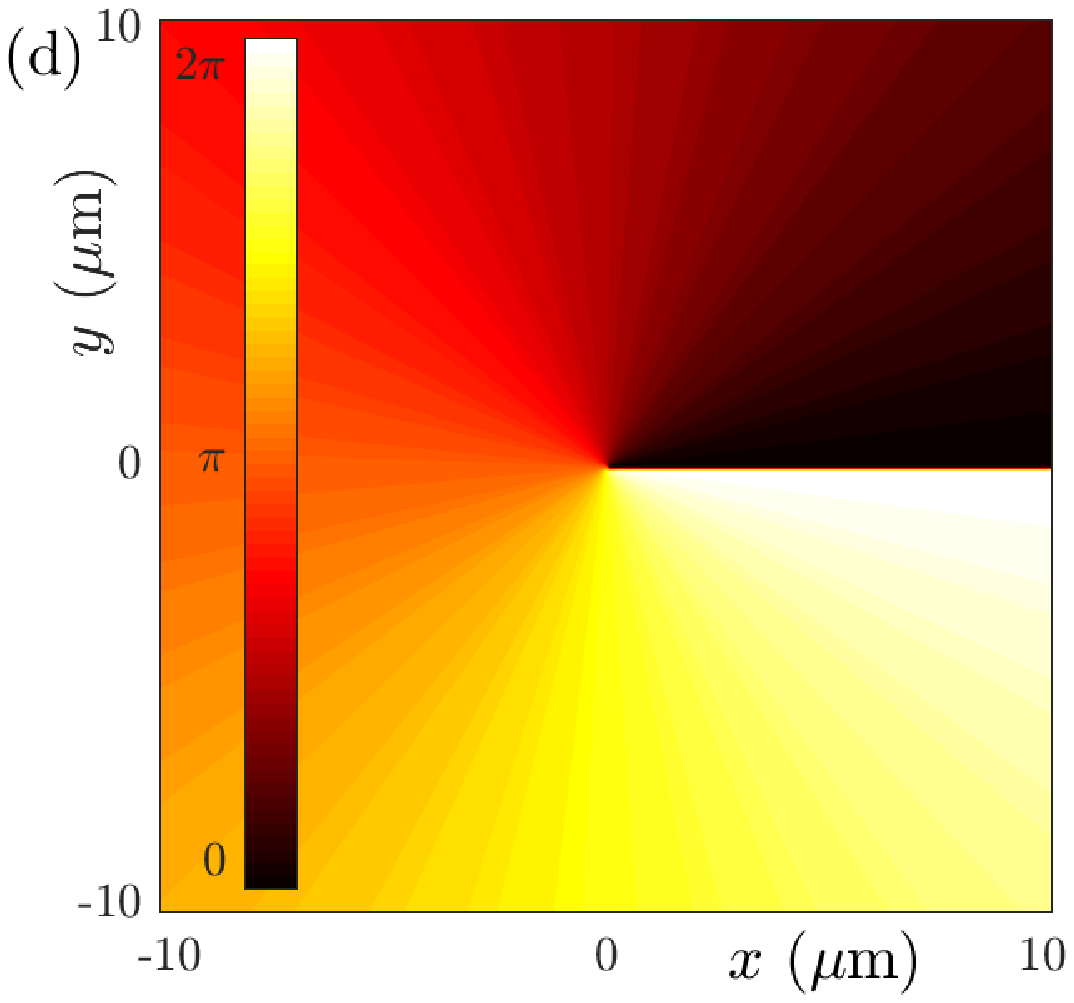}
\end{center}
\caption{\label{Fig1new} {\it Standard OVSs.} (a) Radial amplitude profiles of OVSs. For $m=1$, (b) amplitude isosurfaces $0.1,0.25,0.5,0.75,0.95\times \sqrt{I_0}$ (from darker to lighter) in spacetime $(t',x,y)$, (c) amplitude and (d) and phase profiles at a transversal plane. In this and the rest of the figures, we have taken typical values $\omega_0= 2.355$ rad/fs ($\lambda=0.8$ $\mu$m), $\delta =-0.01$ $\mu$m$^{-1}$, and for the material medium $k_0=11.414$ $\mu$m$^{-1}$ ($n_0=1.453$, yielding a phase velocity $v_f=\omega_0/(k_0+\delta)= 0.206$ $\mu$m/fs), $k'_0= 4.894$ fs/$\mu$m (group velocity $v_g=1/k'_0=0.204$ $\mu$m/fs), and $k_0^{\prime\prime}=-2.893$ fs$^2/\mu$m.}
\end{figure*}

\begin{figure}[b]
\begin{center}
\includegraphics*[height=4cm]{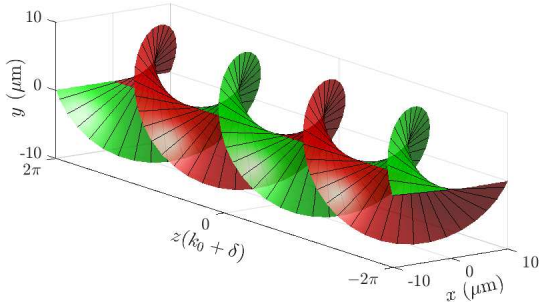}
\end{center}
\caption{\label{Fig2new} {\it Standard OVSs.} Helical wave fronts $m\varphi +(k_0+ \delta)z- \omega_0 t=\mbox{const.}$ for $m=1$ at $t=0$ with $\mbox{const} =0$ and $\pi$.}
\end{figure}

With the trial solution $u=b(\rho)e^{im\varphi} e^{-i\zeta}$, where $\rho=\sqrt{\xi^2+\eta^2}$ and $\varphi = \tan^{-1}(\eta/\xi)$ are polar coordinates in the transversal plane, and $m=\pm 1,\pm 2,\dots$, the radial profile $b(\rho)$ must satisfy
\begin{equation}\label{VS}
\frac{d^2b}{d\rho^2} + \frac{1}{\rho}\frac{db}{d\rho} - \frac{m^2}{\rho^2} b +2b-2b^3 =0\,,
\end{equation}
with boundary conditions $b = C\rho^{|m|}$ for $\rho\rightarrow 0$ and $b \rightarrow 1$ for $\rho\rightarrow \infty$. Solutions to the above boundary problem exist for discrete values $C_1\simeq 0.824754$ for $|m|=1$, $C_2\simeq 0.306198$ for $m=2$, etc. If high accuracy is not required, the radial profiles fit accurately to the expressions $b(\rho) \simeq [\tanh (C_{|m|}^{1/|m|} \rho) ]^{|m|}$. For further comparison, Fig. \ref{Fig1new} shows the normalized radial profile, the amplitude structure in the $(t',x,y)$ space in the form of a cylinder oriented along the $t'$ axis, and transversal amplitude and phase profiles. It is also convenient to recall the helical structure of the phase fronts $m\varphi +(k_0+ \delta)z-\omega_0 t=\mbox{const.}$ in space $(x,y,z)$, as depicted in Fig. \ref{Fig2new} for $m=1$. Phase fronts with any phase between $0$ and $2\pi$ intersect at the line singularity along the $z$-axis.

\begin{figure*}[!]
\begin{center}
\includegraphics*[height=4.2cm]{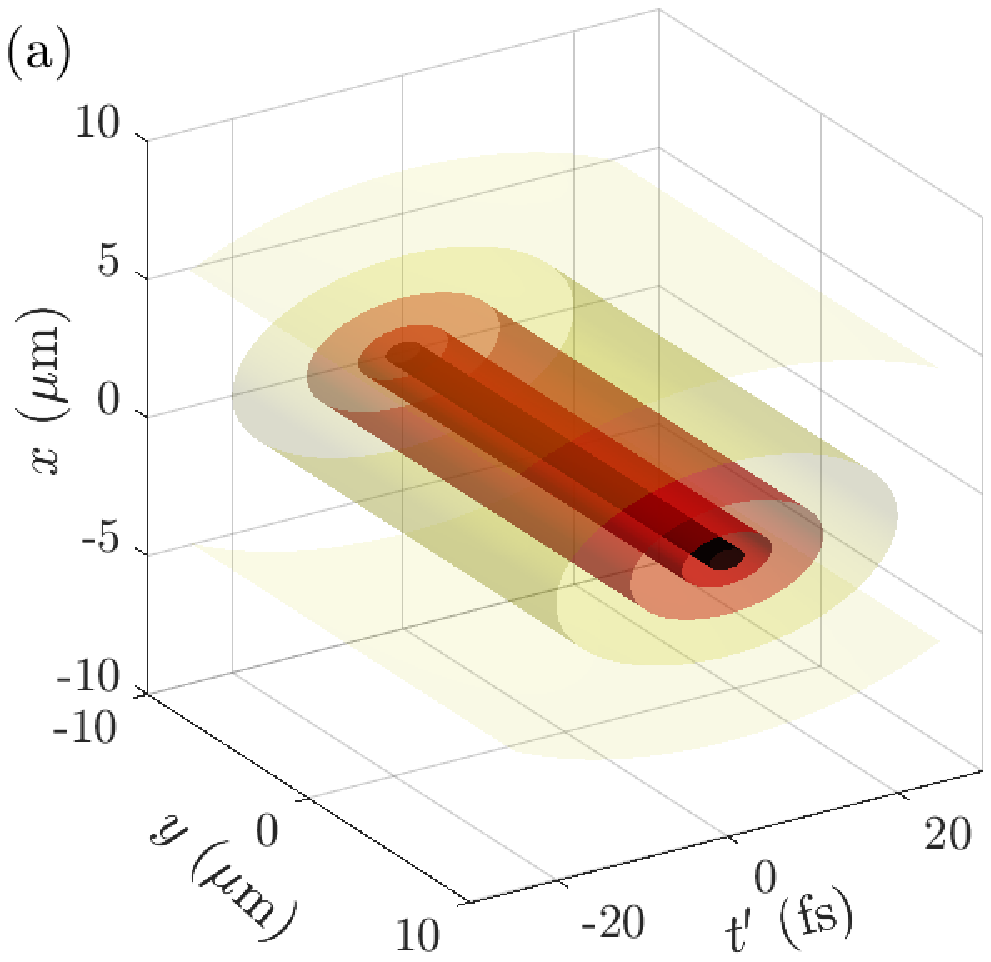}\hspace{0.3cm}\includegraphics*[height=4.2cm]{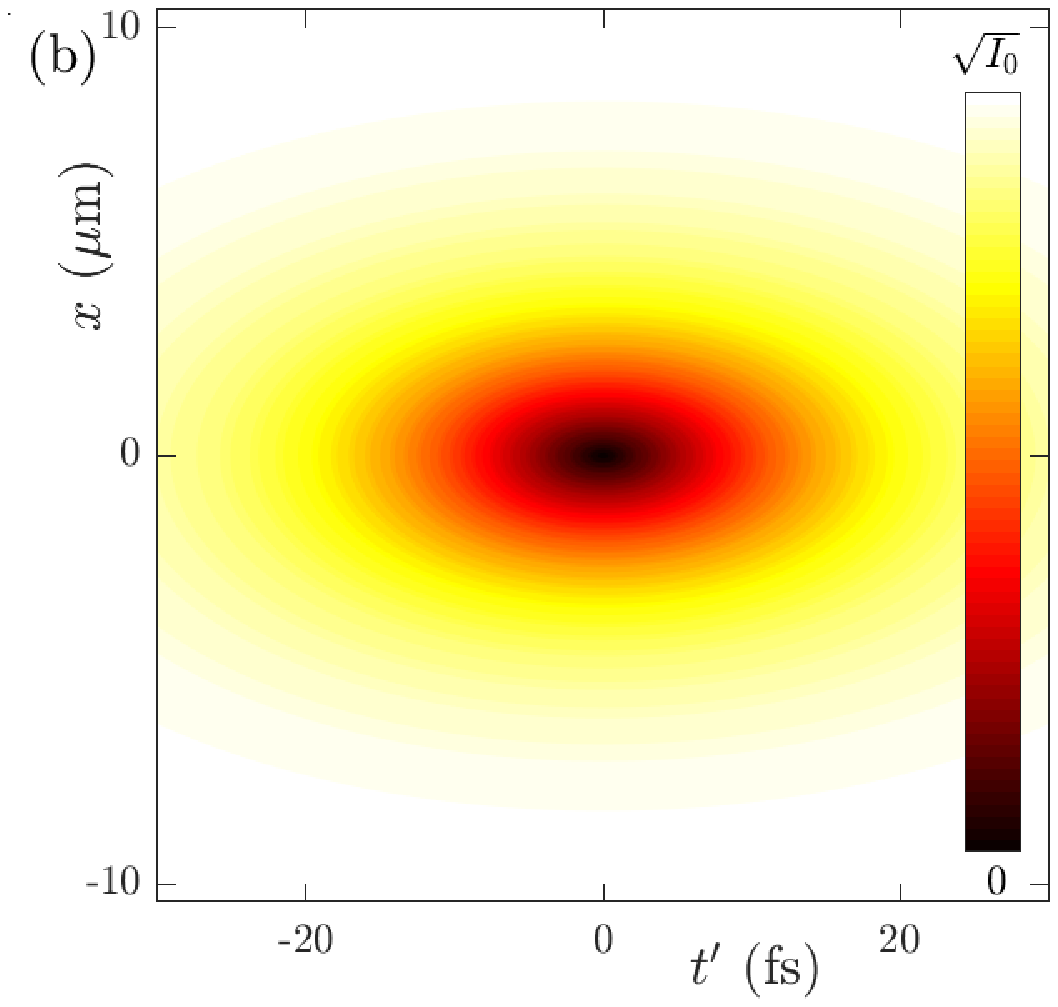}\hspace{0.3cm}\includegraphics*[height=4.2cm]{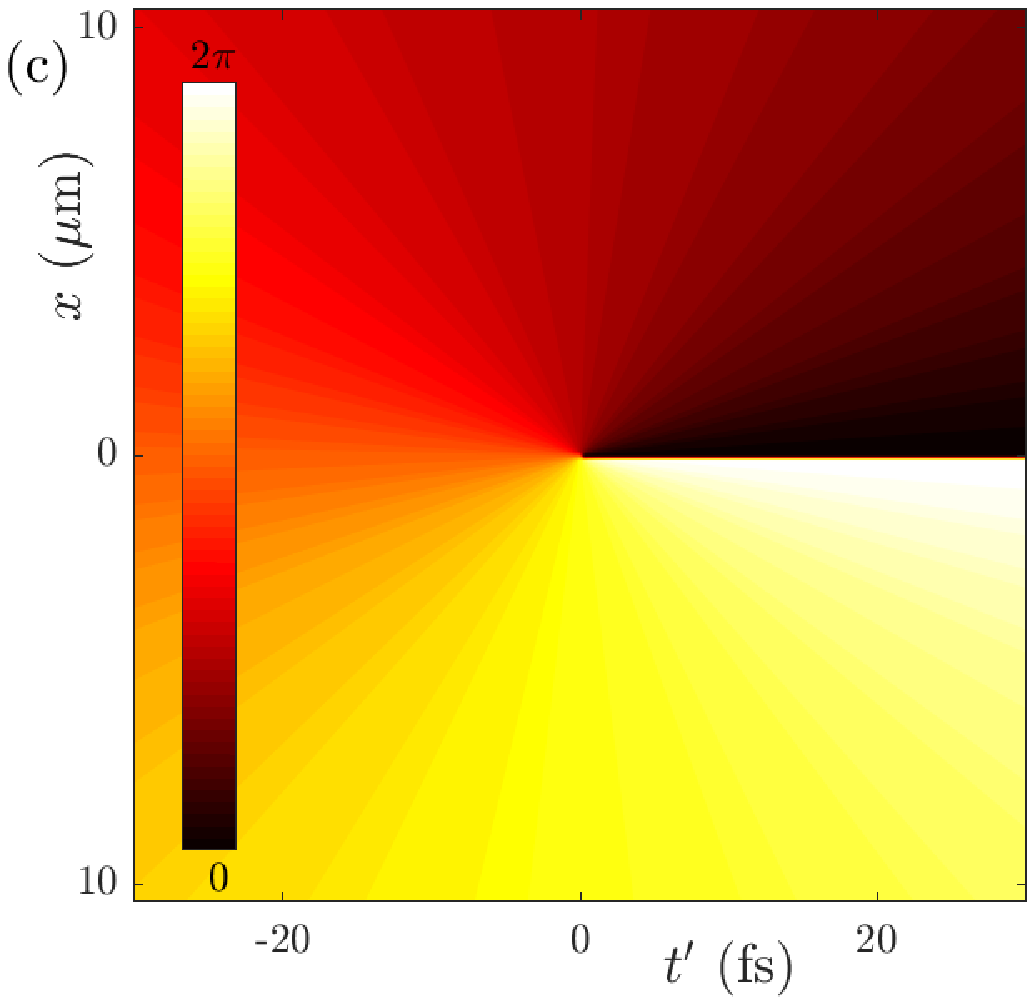}
\end{center}
\caption{\label{Fig3new}  {\it Transversal STOV solitons.} For $m=1$, (a) amplitude isosurfaces $0.1,0.25,0.5,0.75,0.95\times \sqrt{I_0}$ in spacetime $(t',x,y)$, (b) and (c) amplitude and phase profiles in sections $y=\mbox{const.}$ Numerical values used are in Fig. \ref{Fig1new} caption.}
\end{figure*}

{\it STOV solitons with transversal phase line singularity.} A direct extension of the above vortex soliton is a STOV soliton with a line phase singularity oriented along a transversal direction, e. g., the $y$ direction, in a medium with anomalous dispersion, $k_0^{\prime\prime}<0$. Instead of a monochromatic uniform field in $t'$, we now assume a uniform field in $y$, i. e., $u=u(t',\xi,\zeta)$. Introducing the normalized time
\begin{equation}
\tau=\sqrt{|\delta|/|k_0^{\prime\prime}|}\,\, t'\,,
\end{equation}
the NLSE in (\ref{NLSE2}) can be written in the same form as (\ref{NLSE2bis}) but with $(\xi,\eta)$ replaced with $(\tau,\xi)$:
\begin{equation}\label{NLSE3}
\partial_\zeta u = \frac{i}{2}\Delta_{\tau,\xi} u - i|u|^2 u\, .
\end{equation}
Introducing polar coordinates
\begin{equation}
\rho=\sqrt{\tau^2+\xi^2}\,, \quad \varphi = \tan^{-1}(\xi/\tau)\,,
\end{equation}
in the spatiotemporal plane $(\tau,\xi)$, the radial profile $b(\rho)$ of a STOV soliton of the form $u=b(\rho)e^{im\varphi} e^{-i\zeta}$
satisfies the same equation as in (\ref{VS}) with the same boundary conditions. These STOV solitons are then mathematically identical to OVSs but in the plane $(\tau,\xi)$. A detector with $(t',x,y)$ resolution in a given transversal plane would record the a darkness of elliptic cylinder shape about the phase line singularity $x=0$, $t'=0$, i. e., the $y$ axis, as shown in Fig. \ref{Fig3new} (a), i. e., a strip of darkness that appears and disappears as time runs. On any particular $y=\mbox{const.}$ section the amplitude has the elliptical shape, as in Fig. \ref{Fig3new} (b), of half duration of $\sqrt{|k_0^{\prime\prime}|/|\delta|}$, and of $x$-radius $1/\sqrt{k_0|\delta|}$ [taking the radius of $b(\rho)$ in Fig. \ref{Fig1new}(a) as unity for simplicity], and carrying the phase singularity at the center, as in Fig. \ref{Fig3new}(c).

\begin{figure}[b]
\begin{center}
\includegraphics*[width=6.7cm]{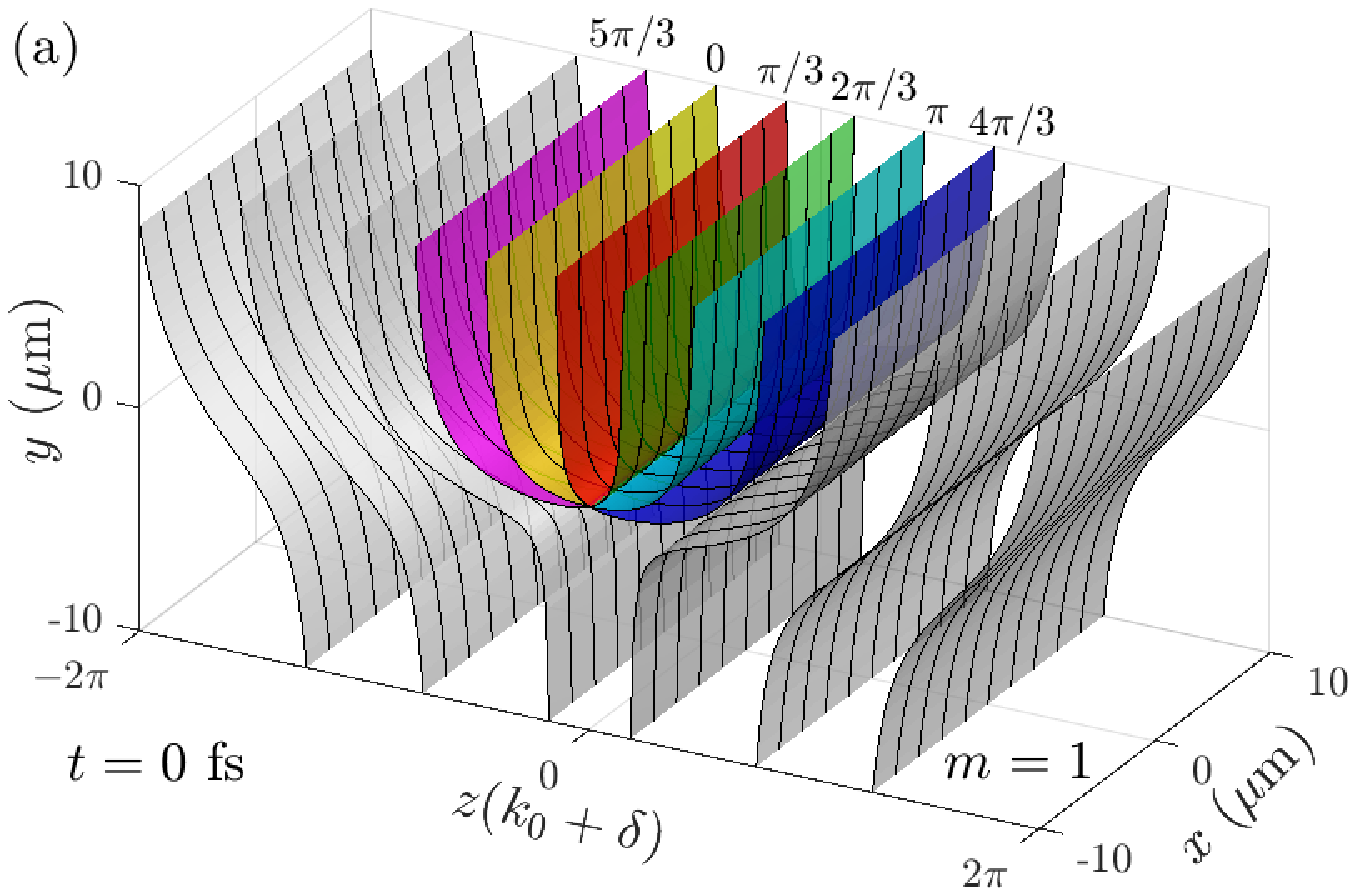}
\includegraphics*[width=6.7cm]{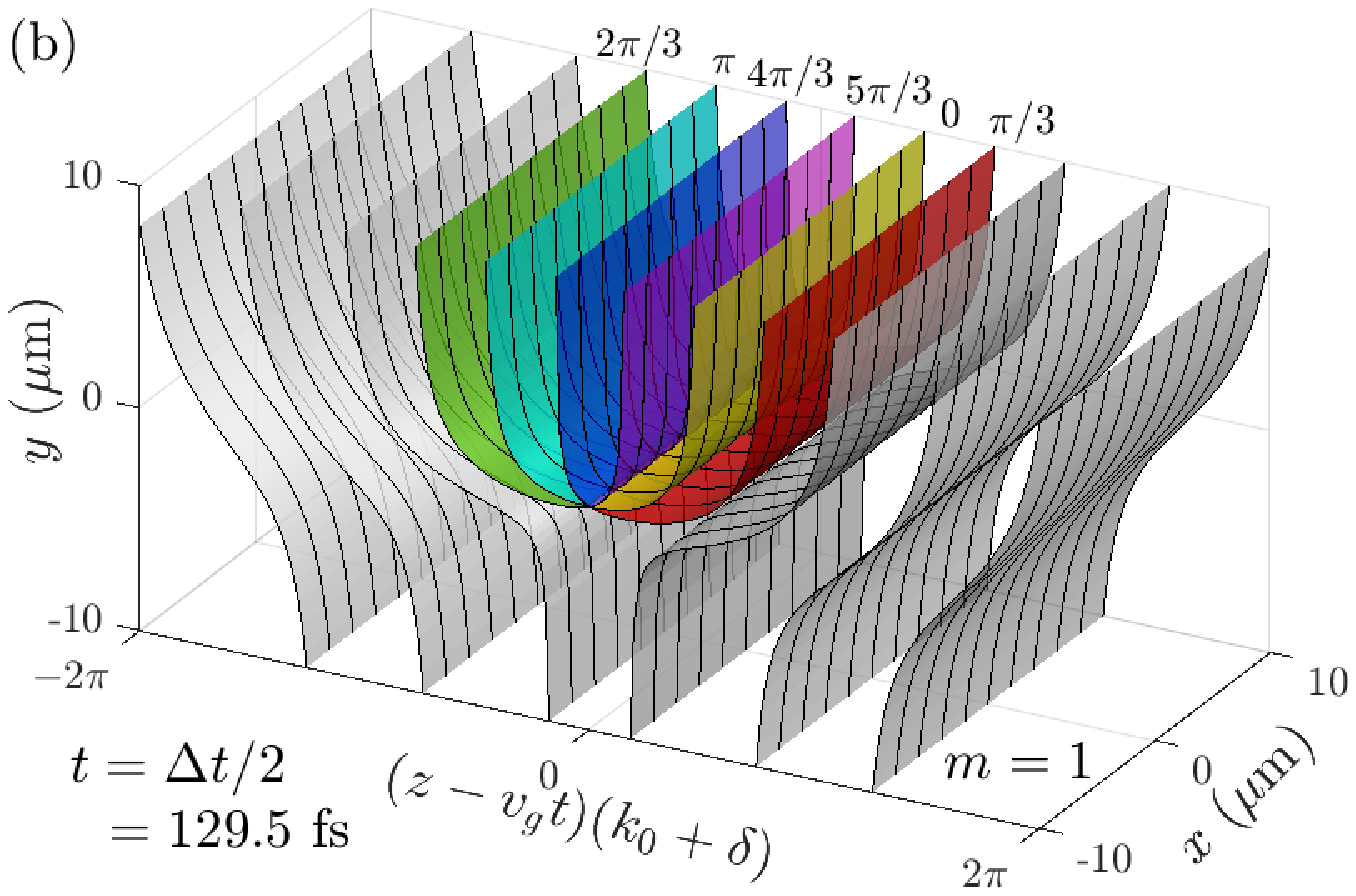}
\end{center}
\caption{\label{Fig4new} {\it Transversal STOV solitons.} (a) Wave fronts $m\varphi +(k_0+ \delta)z-\omega_0 t=\mbox{const.}$ in real space $(x,y,z)$ at $t=0$ with a few indicated values of the constant phase, and $m=1$. At the $y$ axis, wave fronts with all phases between $0$ and $2\pi$ intersect. (b) the same as in (a) but at half a period of wavefront rotation, $t=\Delta t/2=129.5$ fs, corresponding to a propagation distance $z=\Delta z/2=26.46$ $\mu$m. Numerical values used are in Fig. \ref{Fig1new} caption.}
\end{figure}

Despite this seemingly close analogy, the three dimensional behavior of STOV solitons differs radically from that of usual vortex solitons. The difference arises from the fact that it is only the envelope $u$ that has been reoriented, but the carrier oscillations $e^{ik_0 z}e^{-i\omega_0 t}$ continue to propagate along the $z$ direction. Seen in real space $(x,y,z)$ at different laboratory times $t$, the intensity of the STOV soliton is an elliptic cylinder of darkness whose axis is parallel to the $y$ axis, of $x$-radius $1/\sqrt{k_0|\delta|}$ and $z$-radius $\sqrt{|k_0^{\prime\prime}|/|\delta|k^{\prime 2}_0}$, that moves rigidly at the group velocity $v_g=1/k'_0$ along the $z$ direction. The accompanying wave fronts $m\varphi + \delta z +k_0z -\omega_0 t=\mbox{const.}$ have nothing to do with the helical wave fronts of standard OVSs. Several wave fronts at fixed time and with different values of the constant phase are plotted in Fig. \ref{Fig4new}(a) with different colors. Phase fronts with all values of the phase between $0$ and $2\pi$ intersect and terminate at the line singularity along the $y$ axis. Other phase fronts farther from the singularity, shown in gray, tend to match the plane wave fronts of the nonlinear plane wave. When time runs these phase fronts accompany the dark elliptical cylinder at $v_g$, but not rigidly. As an effect of the different phase and group velocities, $v_p=\omega_0/(k_0+\delta)$ and $v_g=1/k'_0$, the phase fronts rotate about the moving singular line, as appreciated in Fig. \ref{Fig4new}(b), repeating themselves with at a time period $\Delta t= (2\pi/k_0)/|v_p-v_g|$, or an axial period $\Delta z =v_g \Delta t = (2\pi/k_0)v_g/|v_p-v_g|$.


{\it STOV solitons with tilted phase line singularity.} Searching for STOV solitons with neither pure longitudinal nor transversal phase line singularity in media with arbitrary dispersion, we try solutions to the NLSE (\ref{NLSE2}) of the form $u=u(t'+\alpha\eta,\xi,\zeta)$. The value $\alpha=0$ corresponds to the above $\eta$-independent STOV solitons, and $\alpha\rightarrow \infty$ to the $t'$-independent, standard OVSs. With general $\alpha$, the ansatz $u=u(t'+\alpha\eta,\xi,\zeta)$ in the NLSE (\ref{NLSE2}) leads to
\begin{equation}\label{NLSE4}
\partial_\zeta u= \frac{i}{2}\left[\left(\alpha^2-\frac{k_0^{\prime\prime}}{|\delta|}\right)\partial^2_{t'}u + \partial^2_\xi u\right] - i|u|^2 u\,.
\end{equation}
For arbitrary real $\alpha$ in media with anomalous dispersion, for $|\alpha|>0$ in media with negligible dispersion, and for $|\alpha|>\sqrt{k_0^{\prime\prime}/|\delta|}$ in media with normal dispersion, we introduce the dimensionless time
\begin{equation}
\gamma= (t' +\alpha \eta)/\sqrt{\alpha^2 - k_0^{\prime\prime}/|\delta|}\,,
\end{equation}
with which the NLSE in (\ref{NLSE4}) simplifies to
\begin{equation}\label{NLSE5}
\partial_\zeta u = \frac{i}{2}\Delta_{\gamma,\xi} u - i|u|^2 u\,.
 \end{equation}
This is again the NLSE (\ref{NLSE2bis}) with $(\xi,\eta)$ replaced with $(\gamma,\xi)$, which admits STOV soliton solutions of the form
$u=b(\rho)e^{im\varphi} e^{-i\zeta}$, where now
\begin{equation}
\rho=\sqrt{\gamma^2+\xi^2}\,, \quad \varphi = \tan^{-1}(\xi/\gamma)
\end{equation}
are polar coordinates in the $(\gamma,\xi)$ plane, with the same spatiotemporal radial profile defined in (\ref{VS}) and depicted in Fig. \ref{Fig1new}(a).

\begin{figure*}[!]
\begin{center}
\includegraphics*[height=4.5cm]{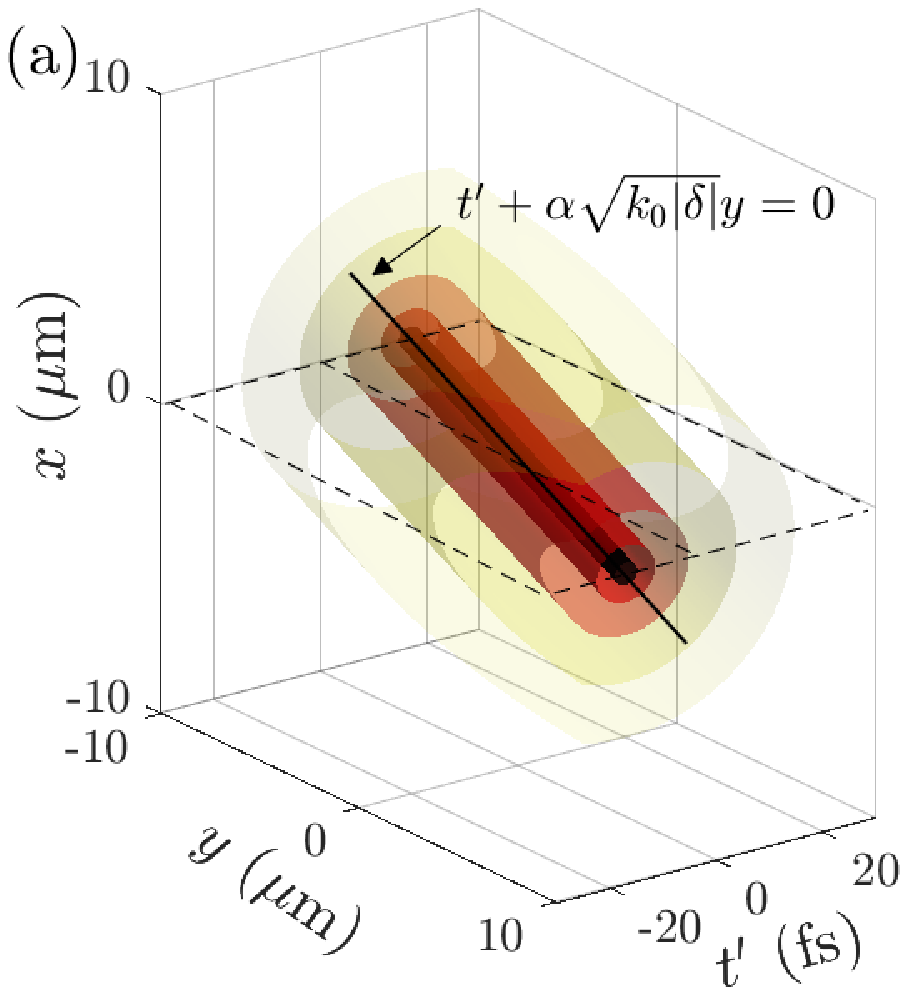}
\includegraphics*[height=3.9cm]{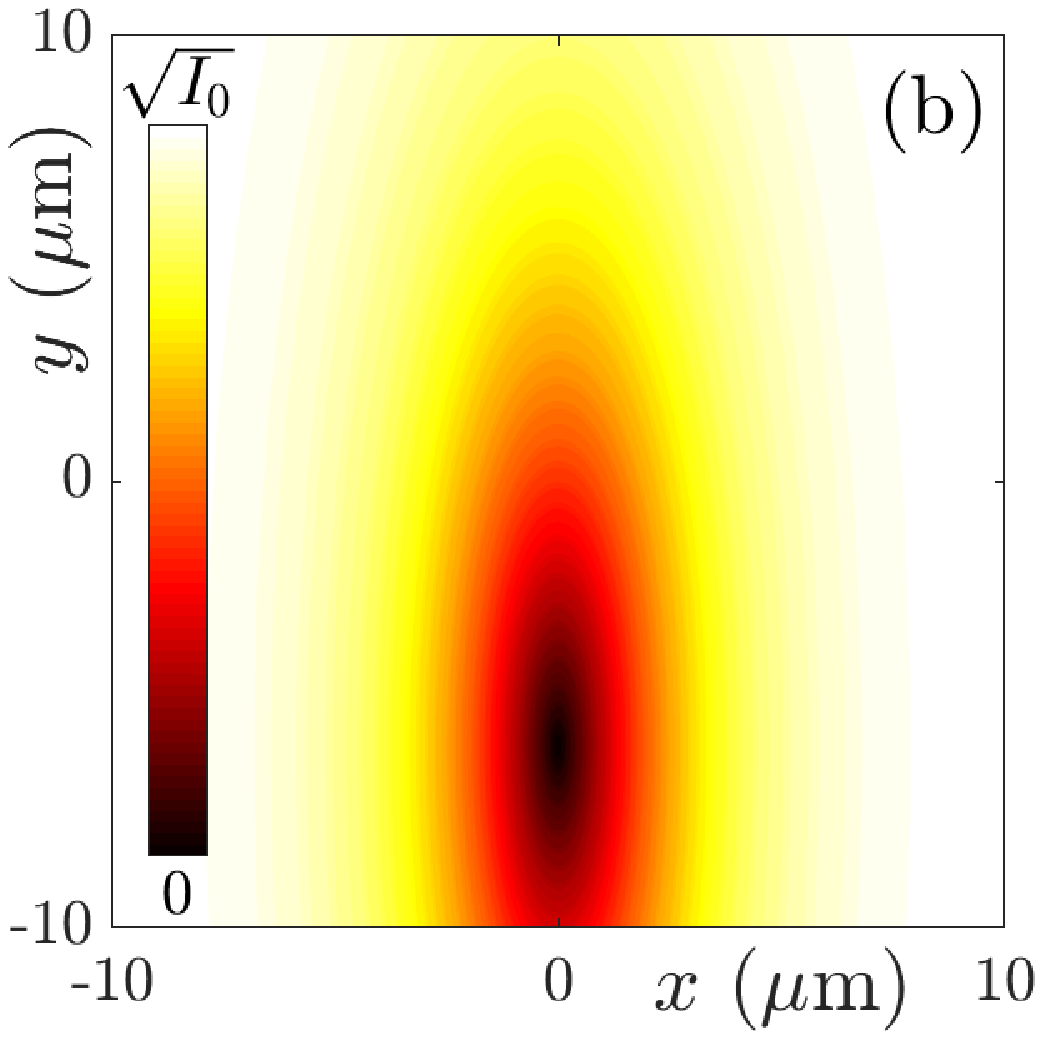}\includegraphics*[height=3.9cm]{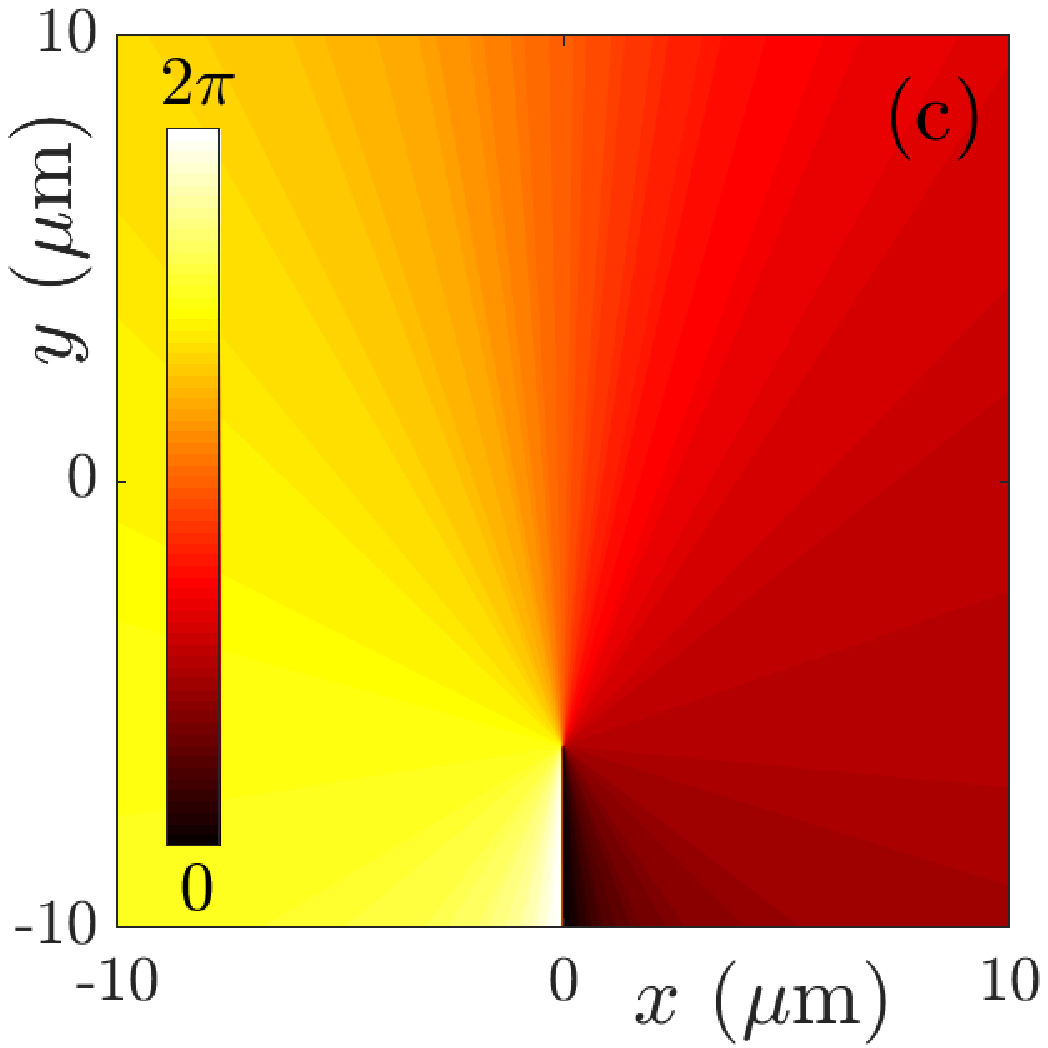}
\includegraphics*[height=3.9cm]{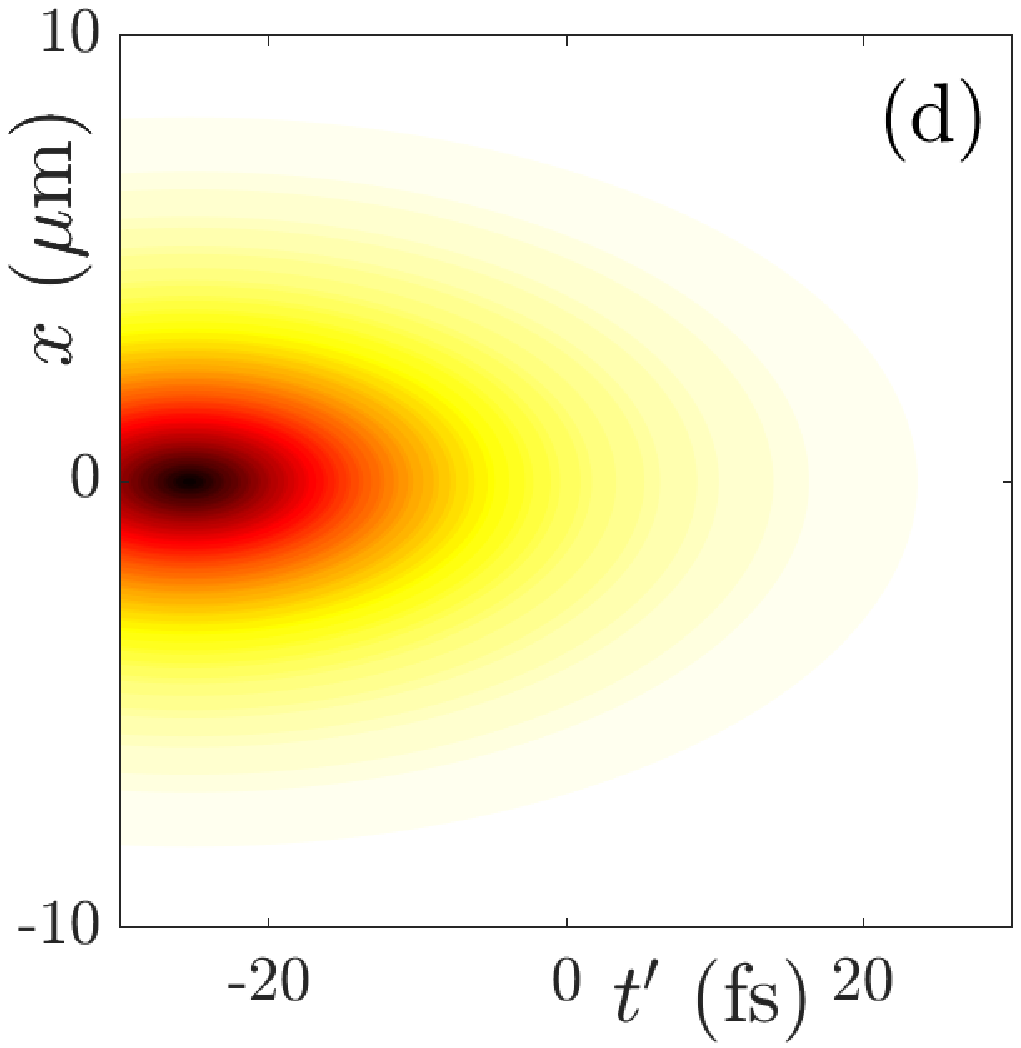}\includegraphics*[height=3.9cm]{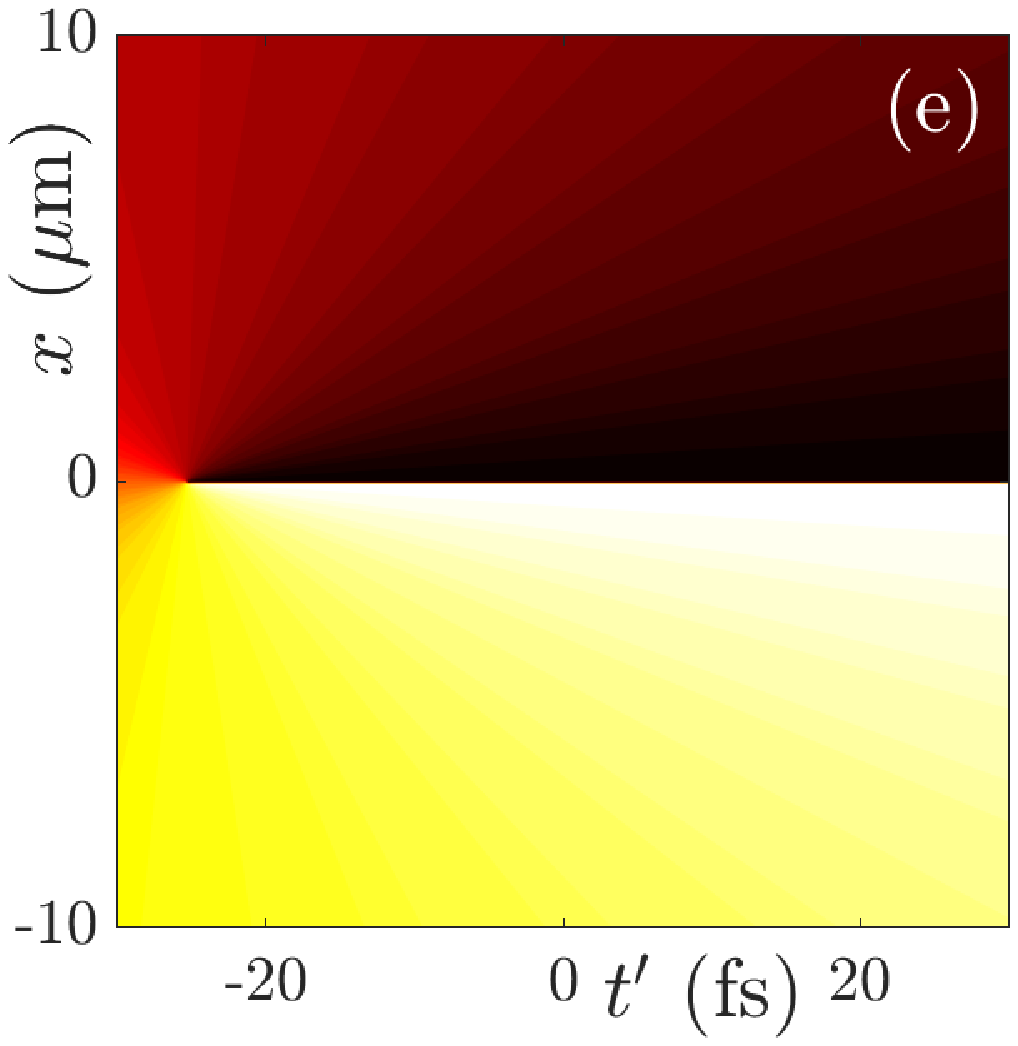}
\end{center}
\caption{\label{Fig5new}  {\it Tilted STOV solitons.} For $m=1$ and $\alpha=5$ fs (a tilt $\beta=71$ deg with respect to the $z$ axis), (a) amplitude isosurfaces $0.1,0.25,0.5,0.75,0.90\times \sqrt{I_0}$ (from darker to lighter) in spacetime $(t',x,y)$. (b) and (c) Amplitude and phase transversal profiles shifted from $(x,y)=(0,0)$ at the time $t'=10$ fs. (d) and (e) Amplitude and phase profiles at section $y=15$ $\mu$m shifted from $(t',x)=(0,0)$. The rest of numerical values used are in Fig. \ref{Fig1new} caption.}
\end{figure*}

Coming back to real variables, a $(t',x,y)$ detector at a transversal plane would record an elliptic cylinder of darkness
surrounding the phase line singularity $x=0$, $t'+\alpha\sqrt{k_0|\delta|}y =0$, as in Figs. \ref{Fig5new}(a). In a transversal plane $(x,y)$ this structure gives an elliptic darkness of $x$ radius $1/\sqrt{k_0|\delta|}$ and $y$-radius $\sqrt{(\alpha^2-k_0^{\prime\prime}/|\delta|)/\alpha^2k_0|\delta|}$ and a phase gradient circulation about the phase singularity that moves along the negative or positive $y$ direction (for positive or negative $\alpha$) at the velocity $-1/\alpha\sqrt{k_0|\delta|}$, as in Figs. \ref{Fig5new} (b) and (c). In the spatiotemporal plane $(t',x)$, the elliptic darkness of duration $\sqrt{\alpha^2- k_0^{\prime\prime}/|\delta|}$, and the phase gradient circulation about the singularity is located at different local times at different $y$-sections, as in Figs. \ref{Fig5new} (d) and (e).

It is also instructive to write STOV solitons explicitly as functions of $(x,y,z)$ at different times $t$. The amplitude is $b(\rho)$ with
\begin{equation}
\rho=\sqrt{k_0|\delta| x^2 + \frac{[z-(t+\alpha\sqrt{k_0|\delta|}y)v_g]^2}{v_g^2(\alpha^2 - k_0^{\prime\prime}/|\delta|)}}\,,
\end{equation}
from where the singular line, $\rho=0$, is $x=0, y= (z-v_g t)/v_g\alpha \sqrt{k_0|\delta|}$. It is then apparent that the amplitude of a STOV soliton has the shape of an elliptical cylinder of darkness tilted an angle given by $\tan\beta = 1/(v_g \alpha \sqrt{k_0|\delta|})$ with respect to the $z$ axis, and that travels undistorted at the group velocity $v_g=1/k'_0$ along the $z$ direction. The $x$ and $y$ sections have the radii given above, and the radius of $z$ sections is $v_g\sqrt{\alpha^2-k_0^{\prime\prime}/|\delta|}$. In media with anomalous dispersion the angle $\beta$ is arbitrary in $[-\pi/2,\pi/2]$. In media with normal dispersion the tilt is limited by $|\beta|< \tan^{-1}(k'_0/\sqrt{k_0 k_0^{\prime\prime}})$, approaching $|\beta|< \pi/2$ if dispersion is negligible.

{\it Discussion and conclusions.} The existence of these solitary vortices raises new questions, some of which can be answered from previous knowledge about OVSs. Being the dynamics of the STOV soliton envelope $u$ governed by mathematically identical equations [the NLSEs (\ref{NLSE3}) and (\ref{NLSE5})] to that for standard OVSs [the NLSE (\ref{NLSE2bis})], all properties that do not involve the carrier oscillations would be the same. For example, STOV solitons would be stable for $|m|=1$, and metastable for $|m|>1$ \cite{DREISCHUH}. Also, the motion of a standard vortex in a non-uniform background is know to be determined by the gradients of the intensity and phase of the background \cite{KIVSHAR2}, and the same is expected to hold for STOV solitons. The interaction between parallel STOV solitons should also be the same as that of OVSs, e. g., two parallel STOV solitons of the same charge rotate around each other and two STOV solitons of opposite charge tend to annihilate. The question that naturally arises is the interaction between non-parallel STOV solitons. In this respect we note that a STOV soliton with negative charge $m<0$ tilted an angle $\beta$ is the same as a STOV soliton with positive charge $m>0$ tilted $\beta+\pi$. Thus rotation or annihilation between parallel STOV solitons can be viewed as different interactions for parallel and antiparallel orientations, which suggests, more generally, that the STOV interactions strongly depends on their mutual orientation.

Finally, in the same way as a standard vortex evolves towards a OVS when introduced in a self-defocusing nonlinear medium \cite{VELCHEV}, the linear STOVs recently generated \cite{HANCOCK,CHONG} would transform into a STOV soliton in the same medium. The generation of these STOV solitons with their new degree of freedom of the orientation of the phase line singularity would open new perspectives in the numerous applications of optical vortices.

The author acknowledges support from Projects of the Spanish Ministerio de Econom\'{\i}a y Competitividad No. MTM2015-63914-P, and No. FIS2017-87360-P.

\end{document}